\documentclass[twocolumn,showpacs,preprintnumbers,amsmath,amssymb]{revtex4}
\usepackage{graphicx}% Include figure files
\usepackage{dcolumn}% Align table columns on decimal point
\usepackage{bm}% bold math
\usepackage{natbib}
\usepackage{amsmath}

\newcommand{\beq}{\begin{equation}}
\newcommand{\eeq}{\end{equation}}

\begin{document}

\title{Experimental Lagrangian structure functions in turbulence}
\author{Jacob Berg$^1$}
\email{jbej@risoe.dtu.dk}
\author{S{\o}ren Ott$^1$}
\author{Jakob Mann$^1$}
\author{Beat L\"{u}thi$^2$}
\affiliation{$^{1}$International collaboration for Turbulence
Research and the Wind Energy Department, Ris{\o} National
Laboratory-DTU, 4000 Roskilde, Denmark \\
$^{2}$International collaboration for Turbulence Research and IFU,
ETH Z\"{u}rich, Switzerland }

\date{\today}
\begin{abstract}
Lagrangian properties obtained from a Particle Tracking
Velocimetry experiment in a turbulent flow at intermediate
Reynolds number are presented. Accurate sampling of particle
trajectories is essential in order to obtain the Lagrangian
structure functions and to measure intermittency at small temporal
scales. The finiteness of the measurement volume can bias the
results significantly. We present a robust way to overcome this
obstacle. Despite no fully developed inertial range we observe
strong intermittency at the scale of dissipation. The multifractal
model is only partially able to reproduce the results.
\end{abstract}

 \maketitle

\section{Introduction}
Turbulent flow still continues to puzzle. Whereas the study of
turbulence in the Eulerian framework has gone through a fruitful
period of discovery the main focus today is on the more subtle
Lagrangian properties of fluid particle behavior. Global issues
such as dispersion of pollutants, cloud dynamics, oceanic food
chain dynamics and a variety of applications ranging from
aerodynamics to combustion need accurate modelling and demand the
latest knowledge on Lagrangian behavior. During the last 10 years
we have experienced a bloom in studies of Lagrangian statistics of
turbulence in fluid flow. Direct Numerical Simulation (DNS)
studies~\cite{yeung:2002,gotoh:2002,biferale:2004,bec:2006a,bec:2006b,yeung:2007}
and laboratory experiments (mainly Particle Tracking Velocimetry
(PTV))~\cite{ott:2000,laporta:2001,mordant:2001,luthi:2005,ayy:2006,bourgoin:2006,berg:2006,luthi:2006b,xu:2006}
have played an important role in revealing the physics governing
the behavior of particles at the smallest scales in a turbulent
flow. Along with this, the theoretical understanding has improved:
the multifractal model~\cite{frisch}, originally introduced in the
Eulerian framework, has now turned into a promising
phenomenological model for Lagrangian observations
~\cite{arneodo:2008}.

In this contribution we present a Lagrangian analysis of small
scale statistical behavior through higher order velocity structure
functions. From a PTV experiment we obtain particle trajectories
and from these we construct structure functions of velocity along
the trajectories. This exercise is common in the field of
turbulence and the above mentioned references all have the
Lagrangian structure functions as the starting point. Studies of
the smallest time scales of the flow has revealed intermittent
behavior. The first signs were observed with DNS. Only within the
last couple of years has it been possible to measure Lagrangian
intermittency in a physical flow with PTV and hence quantitatively
describe the extreme statistics present in the Lagrangian
data~\cite{mordant:2001,mordant:2004b,biferale:2004,xu:2006,xu:2006b}.

The joint work by the International Collaboration of Turbulence
presented in ~\citet{arneodo:2008} showed that the big picture is
the same, whether you use DNS or PTV. DNS and physical flows does,
however, have both quantitative and qualitative differences. DNS
has the disadvantage that at present, due to limited computer
power, the largest structures in the flow can only be simulated a
few times, and hence the statistics becomes very poor on large
scales. As we will argue in this paper, this could have an
influence on even the smallest structures in the flow. In contrast
we can do many independent realizations of the flow with PTV and
make long runs so that all scales are well resolved and
statistically well represented in the many ensembles.
Unfortunately, measuring in a finite volume of the flow can bias
the results significantly: fast particles will leave the volume
early and hence statistics for long times are primarily based on
slow particles. These differences along with a few more should not
be neglected when analyzing data since wrong conclusions could
then easily be made. These issues are the major motivation behind
this paper. We will present a thoroughly way through the jungle of
random errors and bias and try to quantify the importance of each.

The paper is structured as follows:  the technique of PTV is
explained and the flow is characterized in
Section~\ref{PTVandFlow}. In Section~\ref{finiteVolume} we present
a robust way to quantify at which scale the finite volume bias
sets in.  We calculate the final structure functions in
Section~\ref{inertialRangeScaling} and finally make a comparison
with the multifractal model.

Central to the paper are the Lagrangian structure functions
$S_p(\tau)$ of order $p$. With the velocity increment along a
trajectory
\begin{equation}
\delta_{\tau} \mathbf{v}(\tau) \equiv
\mathbf{v}(t+\tau)-\mathbf{v}(t),
\end{equation}
we define a structure function of order $p$ as
\begin{equation}
S_p(\tau)=\langle |\mathbf{e} \cdot \delta_{\tau}
\mathbf{v}(\tau)|^p \rangle,
\end{equation}
with $\langle \cdot \rangle$ denoting ensemble averaging as well
as averaging over all possible directions of the axes, i.e. the
random unit vector $\mathbf{e}$:
\begin{eqnarray}
S_p(\tau) &=&\langle |\delta_{\tau}\mathbf{v}(\tau) |^p
|\cos(\delta_{\tau} \mathbf{v}(\tau), \mathbf{e})|^p\rangle \\
\nonumber &=&\langle |\delta_{\tau} \mathbf{v}(\tau) |^p \rangle
\langle |\cos(\delta_{\tau} \mathbf{v}(\tau),
\mathbf{e})|^p\rangle
\\ \nonumber &=& \frac{1}{p+1} \langle |\delta_{\tau} \mathbf{v}(\tau)|^p
\rangle, \label{eq:SpDef}
\end{eqnarray}
where we have used
\begin{equation}
\int_{0}^{\pi} d\theta \sin \theta |\cos^p\theta| = \frac{2}{1+p}.
\end{equation}

We thus present $S_p(\tau)$ in a way independent of any particular
choice of coordinate system since we have included all possible
rotations around the center of some spherical volume. With this
definition the ensemble is close to isotropic. For the flow
studied in this paper isotropy is actually only strictly present
in the center of the tank. The use of this ensemble does have
several advantages. We can compare with other experiments and DNS
regardless of any degree of anisotropy and hence move closer
towards a general understanding of any universal behavior
regardless of anisotropy; deviations from theories such as the
multifractal model will not be explained by the presence of
anisotropy. Since isotropy is a key element in the foundation of
the multifractal model any such analysis must therefore refer to
isotropic behavior and hence isotropic ensembles of data. Attempts
to develop a framework for anisotropy in turbulence has, however,
been done~\cite{mann:1999,biferale:2005d}.

The time scales relevant for particle motion range from the
viscous scale $\tau_{\eta}$ (the Kolmogorov time scale) to the
integral time scale $T_L$. The Reynolds number, measuring the
strength of the turbulence, scales like $Re_{\lambda}\sim
T_L/\tau_{\eta}$. Two-time particle statistics, where the time lag
$\tau=t_1-t_2$ is less than the integral time scale, but larger
than the Kolmogorov time scale $\tau_{\eta}$, are said to be in
the inertial subrange. From an experimentalist's point of view
this means that Lagrangian inertial subrange scaling is very
difficult to obtain compared to Eulerian statistics where the size
of the inertial subrange grows as $Re_{\lambda}^{3/2}$.

K41 similarity scaling predicts that for time lags in the inertial
range $S_p(\tau) \sim \varepsilon^{p/2} \tau^{p/2},$ where
$\varepsilon$ is the mean kinetic energy dissipation. The lack of
similarity introduces corrections to K41 similarity scaling. These
are due to intermittent events. The multifractal
model~\cite{frisch} is today the most used model of intermittency.
The lack of an inertial subrange of $S_p(\tau)$ observed in
experiments and DNS indicates that the Reynolds number is not the
crucial factor for observing intermittency and data from a range
of Reynolds numbers have also been observed to follow each other
closely~\cite{arneodo:2008}.

\section{Particle Tracking Velocimetry} \label{PTVandFlow}
\subsection{Experimental Setup}
\begin{figure}[htp]
\includegraphics[width=0.8\columnwidth]{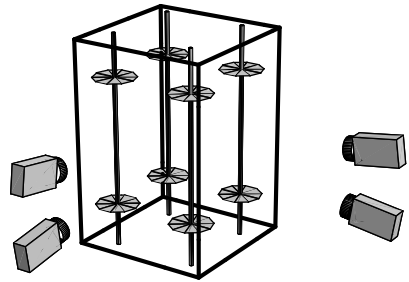}\vspace{-2cm}
\caption{{\footnotesize Experimental setup}} \label{fig:exp}
\end{figure}

We have performed a Particle Tracking Velocimetry (PTV) experiment
in an intermediate Reynolds number turbulent flow. PTV is an
experimental method suitable for obtaining Lagrangian statistics
in turbulent flows. Lagrangian trajectories of fluid particles in
water are obtained by tracking neutrally buoyant particles in
space and time. The flow is generated by eight rotating
propellers, which change their rotational direction in fixed
intervals in order to suppress a mean flow, placed in the corners
of a tank with dimensions $32\times32\times50\mathrm{cm}^3$ (see
Fig~\ref{fig:exp}).
The data acquisition system consists of four commercial CCD
cameras with a maximum frame rate of $50\mathrm{Hz}$ at $1000
\times 1000$ pixels. The measurement volume covers roughly $1000
\mathrm{cm}^3$. We use polystyrene particles with size
$\sim400\mathrm{\mu m}$ and density very close to
$1\mathrm{g}/\mathrm{cm}^3$. We record $\mathcal{O}(800)$
particles at each time step with an accuracy in the estimation of
the particle position of $0.05$ pixels corresponding to a standard
deviation $\sigma_{jitter}\sim 10\mathrm{ \mu m}.$ The particles
are illuminated by a $250\mathrm{W}$ flash lamp.

The Stokes number $St=\tau_R/\tau_{\eta}$ measures the ratio
between the relaxation time, $\tau_R$, of particle motion relative
to the fluid and the Kolmogorov time scale, $\tau_{\eta}$. Here
$\tau_R=(1/18) (\rho_p/\rho_f) (d_p^2/\nu)$ where $\rho_f$ is the
density of the fluid and $\rho_p$ and $d_p$ are the density and
the size of the particles respectively. We get $St=0.01$ and thus
much less than one. The particles can therefore be treated as
passive tracers in the flow.

The mathematical algorithms for translating two dimensional image
coordinates from the four camera chips into a full set of three
dimensional trajectories in time involve several crucial steps:
fitting 2d gaussian profiles to the 2d images, stereo matching
(line of sight crossings) with a two media (water-air) optical
model and construction of 3d trajectories in time by using a
kinematic principle of minimum change in acceleration
\cite{willneff:2003,ouellette:2006}.

If a particle can not be observed from at least three cameras the
linking ends. Most of the the time this happens because the
particles shade for each other. The higher the seeding density of
particles the higher the risk of shadowing. Since the seeding is
relatively high we obtain shorter tracks than we would have in the
case of only a few particles in the tank. Since the particles most
often only disappear from one or two cameras for a few time steps,
a new track starts when the particle is again in view from at
least three cameras. The track is hence broken into smaller
segments. Through kinematic prediction we are able to connect the
broken track segments into longer tracks. We use the method of
\citet{haitao:2008}. The result is satisfactory with a substantial
increase in the mean length of the tracks.

The flow characteristics are presented in Table~\ref{table:flow}.
With $\tau_{\eta}=0.08\mathrm{s}$ and a recording frequency at
$50\mathrm{Hz}$ the temporal resolution is
$\sim4\mathrm{frames}/\tau_{\eta}$. The mean flow is axisymmetric
with a significant vertical straining on the largest scales and no
 significant differences from the flow reported in
\citep{berg:2006,luthi:2006b}, where properties of the mean flow
can be found.

\begin{figure}[htp]
\begin{center}
\includegraphics[width=\columnwidth]{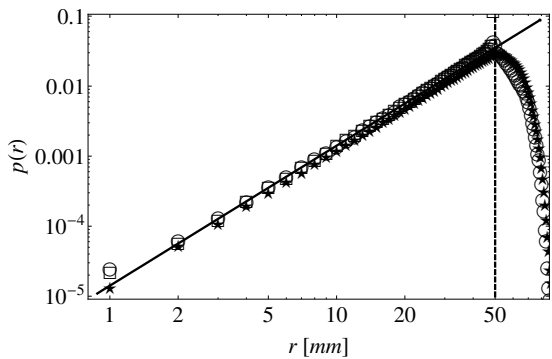}
\caption{{\footnotesize Pdf $p(r)$ of the radial position of
particles in the measuring volume ($\star$ symbol). The symbols
$\Box$ and $\circ$ represent respectively particles starting and
ending a trajectory. The black solid line has a slope equal to two
and therefore represent the uniform distribution.}}
\label{fig:distribution}
\end{center}
\end{figure}

We choose a coordinate system centered approximately in the center
of the tank where the velocity standard deviation $\sigma_u$ has a
global minimum. The radial distribution of particles in the
measuring volume is presented by the stars in
Figure~\ref{fig:distribution}. We see that for distances from the
center less than $50\mathrm{mm}$, represented by the vertical
dashed line, the particles are uniformly distributed. We therefore
choose a ball, $B$, with same the center and a radius of
$50\mathrm{mm}$, as the volume for all future studies in this
paper.
In addition we see that both the start and end position of
trajectories are also uniformly distributed within the ball, $B$.
This means that tracking failure is independent of position.
Trajectories may move in and out of $B$ but only positions inside
the sub-volume are considered in the data analysis.

\begin{figure}[htp]
\begin{center}
\includegraphics[width=\columnwidth]{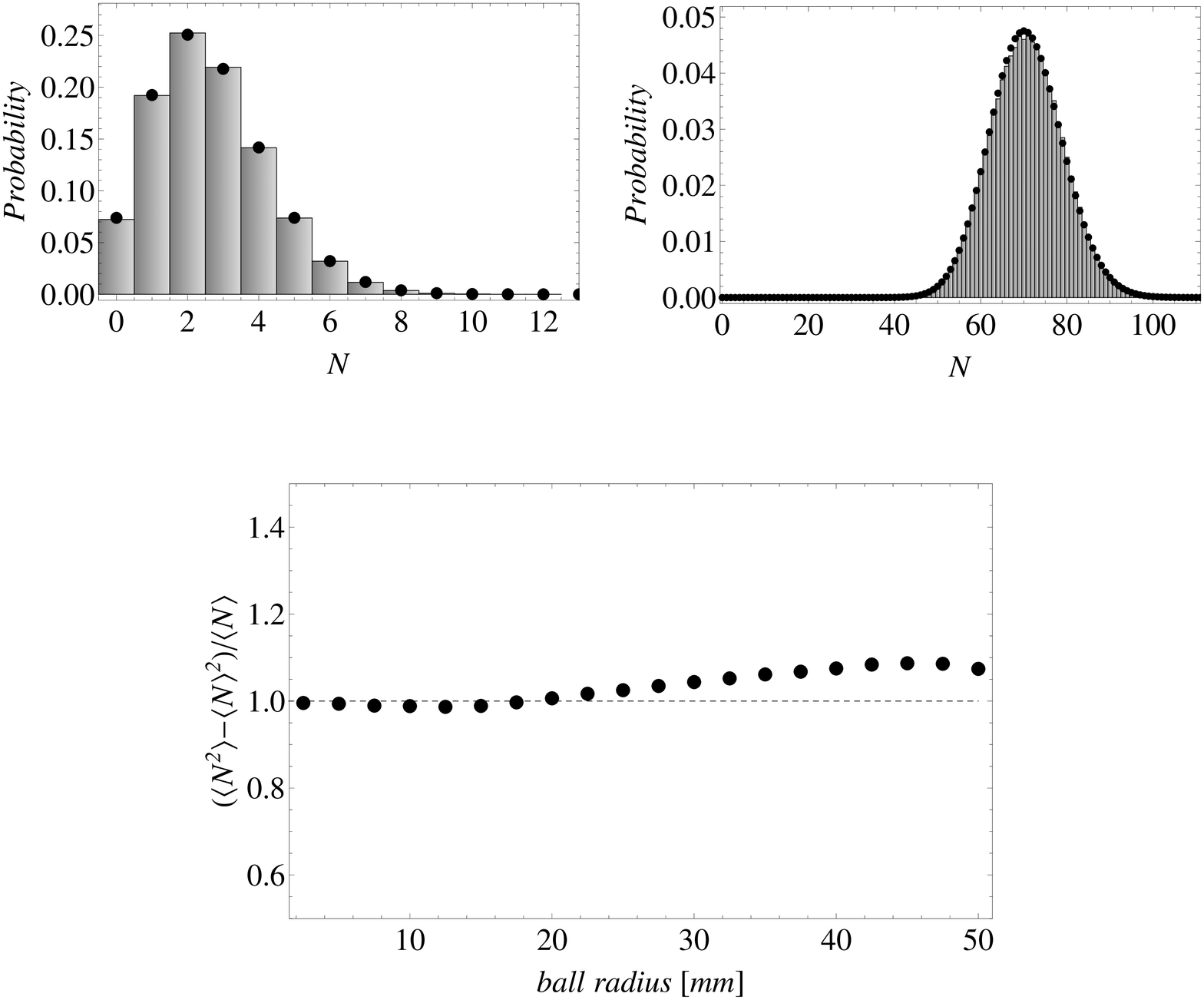}
\caption{{\footnotesize {\em Top:} PDF histogram of number of
particles within randomly positioned balls of radius $10\mbox{
mm}$ ({\em left}) and $30\mbox{ mm}$ ({\em right}). The dots
represent a Poisson distribution. {\em Bottom:} Variation of the
ratio $(\langle N^2\rangle-\langle N \rangle^2)/\langle N \rangle$
as function of ball radius. The horizontal line represent the
Poisson value of unity.}} \label{fig:distribution2}
\end{center}
\end{figure}

To see whether the particles are statistically independent we
check the data against a Poisson distribution. For every tenth
frame we place 100 balls of varying radius randomly within the
flow and count the number of particles inside. The result is
presented in Figure~\ref{fig:distribution2}. The top figures show
two examples; one for particles inside balls of radius $10\mbox{
mm}$ and one where the ball radius is $30\mbox{ mm}$. From the
bottom figure we conclude that the particles obey a Poisson
statistic within 8\% which is a bit better than the PTV experiment
presented in~\cite{berg:2005}. This means that the particles
either cluster or repel each other.

\begin{center}
\begin{table}[t]
\begin{tabular}{|c|c|c|c|c|c|c|}
  \hline
  $\sigma_u$ & $\varepsilon$ & $\eta$ & $L$ & $\tau_{\eta}$ & $T_E$ & $Re_{\lambda}$ \\
  \hline\hline
  $20.00\mathrm{mm/s}$ & $145\mathrm{mm^2/s^3}$ & $0.26\mathrm{mm}$ & $55.17\mathrm{mm}$ & $0.08\mathrm{s}$ & $2.76\mathrm{s}$ & 136 \\
  \hline
\end{tabular}
\caption{{\footnotesize Turbulence characteristics: $\varepsilon$
is the mean kinetic energy dissipation,
$\eta\equiv(\nu^3/\varepsilon)^{1/4}$ is the Kolmogorov length
scale with the kinematic viscosity
$\nu=0.89\mathrm{mm}^2/\mathrm{s}$ of water. $\tau_{\eta}\equiv
(\nu/\varepsilon)^{1/2}$ is the Kolmogorov time scale and
$\sigma_u^2=\frac{1}{3}(\sigma_{u_x}^2+\sigma_{u_y}^2+\sigma_{u_z}^2)$
is the standard deviation of velocity. The integral length scale
is defined as $L=\sigma_u^3/\varepsilon$ while $T_E$ is the eddy
turnover time $T_E=L/\sigma_u$. The Reynolds number is defined as
$Re_{\lambda}=\frac{\lambda \sigma_u}{\nu}$ with the Taylor micro
scale $\lambda=\sqrt{\frac{15 \nu \sigma_u^2}{\varepsilon}}$}}
\label{table:flow}
\end{table}
\end{center}

The database of trajectories is compiled from $73$ runs performed
under identical conditions. Each run consists of $10000$
consecutive frames. After the recording of a run the system was
paused for three minutes before a new run was recorded. We
therefore consider the $73$ individual runs to be statistically
independent. Throughout the paper error bars will therefore be
calculated as statistical standard errors of the mean.

\subsection{Binomial filtering} \label{binomialFiltering}
Even though we know the position error $\epsilon_{pos}$ to be very
small, we choose to filter the data. We choose a binomial filter
which has the convenient property of compact support. Using a
binomial filter instead of a conventional Gaussian, does not seem
to have a large effect (not shown). The weights $w_k$ in a
binomial filter of length $N$ is given by
\begin{equation}
w_k =2^{1-N} \binom{N-1}{k} \hspace{0.5cm} k=0,1,...,N-1
\end{equation}
The width of the filter is $\sigma_{filter}=\Delta t
\sqrt{N-1}/2$. We apply the binomial filter on the position
measurements treating each dimension separately. The velocity and
acceleration are then simply given by finite differences
\begin{eqnarray}
\tilde{v}_i(t+\Delta t/2) &=& \frac{\tilde{x}_i(t+\Delta
t)-\tilde{x}_i(t)}{\Delta t} \\ \nonumber \tilde{a}_i(t) &=&
\frac{\tilde{x}_i(t+\Delta
t)-2\tilde{x}_i(t)+\tilde{x}_i(t-\Delta t)}{\Delta t^2}, \\
\nonumber
\end{eqnarray}
where $\tilde{\cdot}$ denotes filtered quantities.

\begin{figure}[htp]
\begin{center}
\includegraphics[width=\columnwidth]{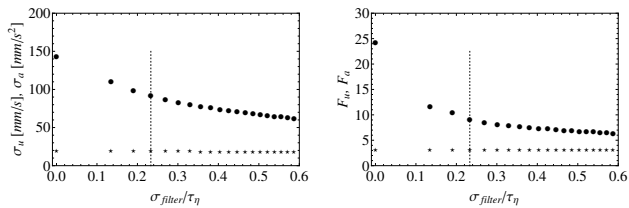}
\caption{{\footnotesize {\em left:} $\sigma_u$ ($\star$) and
$\sigma_a$ ($\bullet$) as functions of filter width
$\sigma_{filter}$ divided by Kolmogorov time scale $\tau_{\eta}$.
{\em right:} Flatness $F_u$ ($\star$) and $F_a$ ($\bullet$). The
dotted vertical lines denote a filter of length $N=4$
corresponding to $0.23 \tau_{\eta}$.}} \label{fig:filter}
\end{center}
\end{figure}

\begin{figure}[h]
\begin{center}
\includegraphics[width=\columnwidth]{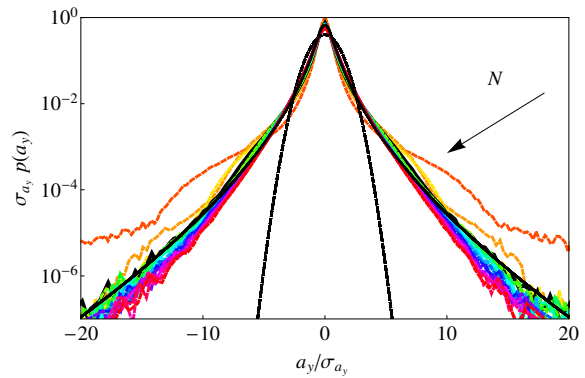}
\caption{{\footnotesize Acceleration pdf $p(a_y)$ for the
component along the axis of forcing in the tank. The pdf is
normalized with the standard deviation of acceleration
$\sigma_{a_y}$ for the same component. The different colors
represent the different filter lengths $N$ ranging from $1$ to
$20$ following the direction of the arrow. The two black curves
are the gaussian distribution and a stretched exponential fitted
to the curve for $N=4$ (only partly visible as the thin black
curve below the fitted stretched exponential).}}
\label{fig:accPDF}
\end{center}
\end{figure}

We inspect filters with a length from $N=1$ (unfiltered) to $N=20$
and look at the standard deviation and flatness of velocity and
acceleration as functions of the filter standard deviation
relative to the Kolmogorov time scale $\tau_{\eta}$. These are
displayed in figure~\ref{fig:filter}. While velocity seems to be
unaltered by the filtering the acceleration is very dependent on
filter length. This is a general problem with measurements of
accelerations.
Both the standard deviation and flatness are functions of the
filter width. The amount of filtering is a trade off between
eliminating noise and eliminating the real signal. The figure does
not give a clear indication of which filter to use. We therefore
also look at the acceleration pdf. Since this is the limiting PDF
for velocity increments $\delta \mathbf{v}(\tau)$ it might seem
important for the outcome of results on Lagrangian structure
functions. In figure~\ref{fig:accPDF} we show the pdf of
acceleration as a function of the filter length.
From $N=4$ the shape of the pdf becomes more or less constant with
$N$. The fat tails of the pdf based on unfiltered data ($N=1$) are
likely to be due to noise and perhaps bad connection of tracks.
Filtering removes the noisy tails and makes it possible to
estimate moments up to $\langle a^8 \rangle$. The critical filter
length for which convergence is achieved is $N=4$. $N=4$
corresponds to a filter width of $0.23 \tau_{\eta},$ i.e. well
below the Kolmogorov scale. It should be emphasized that choosing
$N=4$ is still somewhat arbitrary since there is no rigorous way
to determine the optimal filter. Unless otherwise stated $N=4$ has
been used in the remainder of the paper.

We have fitted a stretched exponential of the form $p(a)=N_f
\exp(-a^2/((1+|a \beta_f/\sigma_f|)^{\gamma_f}) \sigma_f^2)$ to
the pdf of scaled acceleration for $N=4$. With $\beta_f=0.65,$
$\sigma_f=0.55$ and $\gamma_f=1.40$ the fit is excellent. The same
functional fit for $p(a)$ was used by \citet{laporta:2001} with
slightly different fit parameters. We can also give an estimate
for the dimensionless constant $a_0$ in the Heisenberg-Yaglom
relation $\langle a_i a_j \rangle = a_0 \varepsilon^{3/2}
\nu^{-1/2} \delta_{ij}$. In principle $a_0$ is not a constant but
a function of Reynolds number. It has been well studied in the
literature and it is closely connected to Lagrangian stochastic
models. \citet{sawford:2003} fits functional forms of $a_0$ as
function of Reynolds number to data obtained from DNS and high
Reynolds number PTV. The values obtained from the PTV data are
approximately $30\%$ larger than for the corresponding DNS data.
In addition strong anisotropy is observed in the PTV data. We get
$a_0=4.5\pm0.3$ for the component averaged acceleration variance
while only $a_0=3.8\pm0.2$ for the component along the axis of
forcing. For the axial component \cite{sawford:2003} suggest
$a_0=6.5(1+134 Re_{\lambda}^{-1} )^{-1}$ giving $a_0=3.3$ for the
present Reynolds number. The construction of our apparatus does,
however, not give us the opportunity to investigate a large span
in Reynolds numbers which would be necessary to verify functional
forms of $a_0$.

As already mentioned the filtering has a large impact on
acceleration. Besides that, there is a slight chance that we might
underestimate accelerations since variations at the smallest
temporal scales simply cannot be resolved. In the PTV data
presented in \cite{sawford:2003,laporta:2001} the ratio between
$\tau_{\eta}$ and sampling frequency is $\sim23\mbox{
frames}/\tau_{\eta}$ (for $Re_{\lambda}\sim 870$) compared to only
$\sim4\mbox{ frames}/\tau_{\eta}$ in our experiment. Their data
is, however, much noisier, so that their chosen filter width of
$\sigma_{filter}/\tau_{\eta}=0.15$ is close to our of $0.23$ for
$N=4$.

The motivation behind filtering the trajectories was to eliminate
noise, i.e. the error, $\epsilon_{pos}$, associated with
determining the position of a particle. Even though
$\epsilon_{pos}$ might be both random, unbiased and uncorrelated
it still contributes to, for example, the Lagrangian structure
functions. To see this we assume that
$x_i(t)=\hat{x}_i(t)+\epsilon_{pos}(t)$. $x_i(t)$ is a measured
component of position on a trajectory while $\hat{x}_i(t)$ is the
{\em true} position. Furthermore we assume $\langle
\epsilon_{pos}(t) \rangle=0$ and $\langle
\epsilon_{pos}(t_1)\epsilon_{pos}(t_2)\rangle = \sigma_{jitter}^2$
if $t_1=t_2$ and zero otherwise.

The correction to the structure functions is now straight forward
to calculate. Since a different number of particle positions are
involved in the calculation as a function of time lag, $\tau$, the
correction becomes a function of the time lag, $\tau$, itself.
With filter $N=4$ we get for the second order measured structure
function $S_2^{m}(\tau)$
\begin{equation}
S_2^{m}(\tau)=S_2(\tau)+\sigma^2_{error}(\tau)
\end{equation}
where $\sigma^2_{error}(\tau)$ is given by
\begin{equation}
\sigma^2_{error}(\tau)=\begin{cases}
\frac{3}{16} \frac{\sigma_{jitter}^2}{\Delta t^2} & \text{for $\tau=\pm \Delta t$}, \\
\frac{7}{16} \frac{\sigma_{jitter}^2}{\Delta t^2} & \text{for $\tau=\pm 2\Delta t$}, \\
\frac{7}{16} \frac{\sigma_{jitter}^2}{\Delta t^2} & \text{for $\tau=\pm 3\Delta t$}, \\
\frac{11}{16} \frac{\sigma_{jitter}^2}{\Delta t^2} & \text{for $\tau=\pm 4\Delta t$}, \\
\frac{5}{16} \frac{\sigma_{jitter}^2}{\Delta t^2} & \text{for $\tau =\pm n\Delta t,$ $n \geq 5$}. \\
\end{cases}
\end{equation}
We find that the correction to the measured higher order structure
functions $S_p^m(\tau)$ is given by
\begin{equation}
S_p^{m}(\tau)=S_p(\tau)+A_p \sigma^2_{error}(\tau)
S_{p-2}^m(\tau),
\end{equation}
with $S_0=1$ for all time lags $\tau$. We have omitted terms of
higher order in $\sigma_{error}$. $A_p$ is given by
\begin{equation}
A_p=\frac{p^2-p}{2}.
\end{equation}
With $\sigma_{jitter}\sim 10 \mu \mathrm{m}$ the corrections
$\sigma^2_{error}(\tau)$ are of order
$10^{-3}\mathrm{mm}^2/\mathrm{s}^2$.

\section{Bias of Lagrangian statistics} \label{finiteVolume}
A practical property of the present experiment is the stationarity
of Eulerian velocity statistics. The Lagrangian statistics are, on
the other hand, not stationary in the measurement volume, $B$.

A particle which enters $B$ will loose kinetic energy during its
travel inside $B$. This reflects the non-uniform forcing in space
in our experiment. On average the particles gain kinetic energy
close to the propellers located outside $B$. During their
subsequent motion the particles lose kinetic energy until they
again come close to the propellers which are constantly spinning.
Thus there is a flux of kinetic energy into $B$. Inside the volume
the kinetic energy is dissipated and hence we have $\frac{1}{2}
\frac{d}{dt} \langle \mathbf{v}^2 \rangle \sim -\varepsilon$
~\cite{ott:2005}. The equation can be derived directly from the
Navier-Stokes equation by assuming global
homogeneity~\cite{mann:1999} which, as already mentioned, is only
approximately true for this experiment.

In DNS the random forcing occurs in fourier space and is hence
globally homogeneous. We therefore have $d\langle v^2
\rangle/dt=0$ and consequently Lagrangian stationarity. However,
most physical flows encountered in nature, as for example the
atmospheric boundary layer, will seen in a finite volume be
Lagrangian non-stationary.

Particles with fast velocity tend to leave the measurement volume
after only a short amount of time. This means that those particles
that stays in the volume for long times often are those with the
smallest velocity. The effect is a bias for long times towards
slow particles. It should be emphasized that this is a systematic
error whereas the Lagrangian non-stationarity is a genuine
property of the flow.

Exactly how one should compensate for the systematic error is an
open question. In this paper we will build on the ideas first
presented by \citet{ott:2005}. Here the Lagrangian structure
functions $S_p(\tau)$ are expressed through the mean Greens
function $G(\mathbf{r},\tau)$ as
\begin{equation}
S_p(\tau)=\int_{\mathcal{R}^3} S_p(\tau|\mathbf{r})
G(\mathbf{r},\tau) d^3 r,\label{eq:greens}
\end{equation}
where $S_p(\tau|\mathbf{r})$ is the conditional Lagrangian
structure function of order $p$ defined as the mean of $|\delta
v|^p$ conditioned on the distance travelled $\mathbf{r}$ after a
time lag $\tau$. The mean Greens function for one-particle
diffusion $G(\mathbf{r},\tau)$ is the probability density of
getting $\mathbf{r}$ after a time lag $\tau$. The relation is
evidently exact. However, when the field of view is limited the
integration on the right hand side is truncated which leads to
systematic errors. It is truncated by a filter $W(r)$ that
expresses the probability that a point $\mathbf{x}_2$ separated a
distance $r$ from another point $\mathbf{x}_1$ lies inside a ball
with radius $R$. Here $\mathbf{x}_1$ is chosen randomly inside the
ball. Assuming homogeneity $W(r)$ is a purely geometric factor of
the distance $|\mathbf{x}_1-\mathbf{x}_2|=r$. It is given by
\begin{equation} W(r)=\begin{cases} (1-\frac{r}{2 R})^2 (1+\frac{r
}{4 R}) & \text{for $r<2 R$}, \\ 0 & \text{for $r>2
R$}.\end{cases}
\end{equation}

If we neglect the finite measurement volume and just average the
velocity differences that we have actually measured, the
experimental structure function becomes
\begin{equation}
S_{p,meas}(\tau)=\frac{\int W(r) S_p(\tau|r) 4 \pi r^2 G(r,\tau)
dr} {\int W(r) 4 \pi r^2 G(r,\tau) dr}. \label{s2W}
\end{equation}
\citeauthor{ott:2005} proposed an improved method where $W(r)$ is
removed from eqn.~\ref{s2W}: each pair is binned with weight
$1/W(r)$. Since $W(2 R)=0,$ pairs with separations very close to
$2R$ should be disregarded since they would otherwise make the
compensation explode. We therefore limit separations to
$2R-\delta$. In the present case $2R=100\mathrm{mm}$ and
$\delta=5\mathrm{mm}$.

Including compensation we can now write eqn.~\ref{s2W} as
\begin{equation}
S_{p,meas}(\tau)=\frac{\int_0^{2R-\delta} S_p(\tau|r) 4 \pi r^2
G(r,\tau) dr}{\int_0^{2R-\delta} 4 \pi r^2 G(r,\tau) dr}.
\label{s2W2}
\end{equation}
\begin{figure}[h]
\begin{center}
\includegraphics[width=\columnwidth]{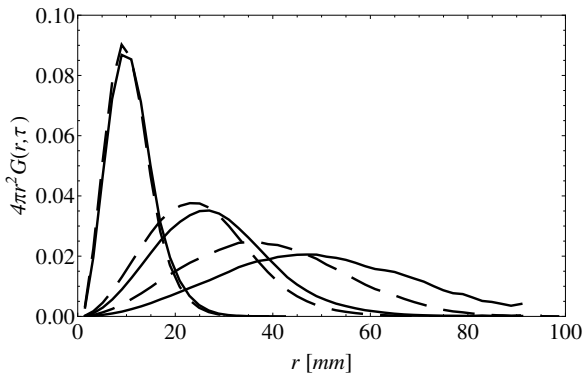}
\caption{{\footnotesize Mean Greens function $G(r,\tau)$ as a
function of $r$ at three different time lags $\tau$. Increasing
towards the right, time lags are $\tau =\{5,13,35\} \tau_{\eta}$
with (solid lines) and without (dashed lines) compensation through
$W(r)$.}}\label{fig:G}
\end{center}
\end{figure}

We will use this framework to estimate an upper time lag below
which Lagrangian structure functions are not biased by the finite
measurement volume. To put it simple we need to figure out whether
or not $2R-\delta$ is enough to cover the support of the
integrands in eqn.~\ref{s2W2}.

In Fig.~\ref{fig:G} we show the mean Greens function $G(r,\tau)$
as a function of $r$. For the small time lag $\tau=5\tau_{\eta}$
 the curves almost collapse indicating that
finite volume effects are almost vanishing. Increasing $\tau$ to
$13\tau_{\eta}$ we see that finite volume effects are significant
as the curves no longer collapse. Both curves, however, go to zero
for large values of $r$, which means that the integrand in
eqn.~\ref{s2W2} converges. This means that for this specific time
lag we can calculate finite volume unbiased structure functions if
we include compensation. For the largest time lag,
$\tau=35\tau_{\eta}$ we see that the compensated $G(r,\tau)$ no
longer converges. This means that even the compensation fails.

\begin{figure}
\begin{center}
\includegraphics[width=1.0\columnwidth]{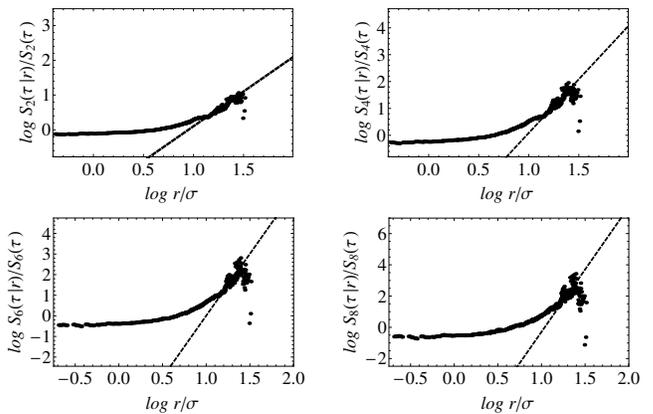}
\end{center}
\caption{$\log S_p(\tau|r)/S_p(\tau)$ as a function of $\log
r/\sigma_x(tau)$ for $\tau \in (15;25)\tau_{\eta}$. The four
panels represents $p=2,4,6,8$ with fits $r^p$ for large
$r/\sigma_x(tau)$.} \label{str}
\end{figure}

\begin{figure}
\begin{center}
\includegraphics[width=1.0\columnwidth]{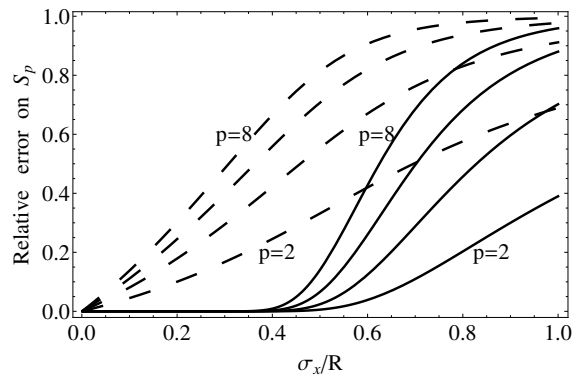}
\end{center}
\caption{Relative error estimates for $S_p$ for $p=2,4,6$ and 8.
Dashed lines are without compensation while solid lines are with
compensation} \label{ErrorPlots}
\end{figure}

We now assume $G(r,\tau)$ to be Gaussian and self similar, that is
$G(r,\tau)=\exp(-(r/\sigma_x(\tau))^2/2)/(2 \pi^{3/2}
\sigma_x(\tau)^3)$. Furthermore we assume that $S_p(\tau|r)\sim
r^p$ for large $r$ and large $\tau$. Figure~\ref{str} shows
log-log plots of $S_p(\tau|r)$ for $p=2,4,6,8$. We observe an
approximately agreement with $S_p(\tau|r)\sim r^p$.
\citet{ott:2005} calculated $S_2$ for a gaussian displacement
process and found that $S_2(\tau|r)\sim r^2$ for large $r$. The
implications for $S_p$ is evident: only the slightest deviation
from zero in $G(r,\tau)$ for large $r$ will cause the integrand in
the numerator of eqn.~\ref{s2W2} not to converge. With the made
assumptions we can calculate the relative error on $S_p$ from the
finite volume.
For the compensated structure function the relative error given by
$1-S_{p,meas}/S_p$. $S_{p,meas}$ can be calculated from
eqn.~\ref{s2W2} while $S_p$ is given by eqn.~\ref{eq:greens}
integrating all the way to infinity. Likewise we can calculate the
error without compensation. In this case $S_{p,meas}$ can be
calculated from eqn.~\ref{s2W}. The results are presented in
Figure~\ref{ErrorPlots} for $p$=2,4,6 and 8. The solid lines
represent the compensated structure functions while the dashed
lines represent the uncompensated. The error is a function of the
non dimensional variable $\sigma_x/R$.

\begin{figure}
\begin{center}
\includegraphics[width=1.0\columnwidth]{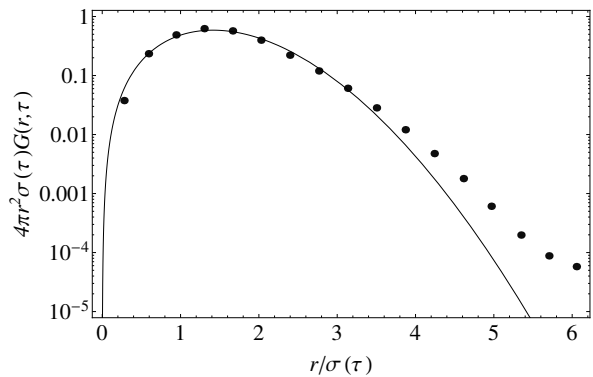}
\end{center}
\caption{$G(r,\tau)$ for $\tau=10\tau_{\eta}$. The full line is a
Gaussian.} \label{fig:Glog}
\end{figure}

It is evident that without compensation we get large, systematic
errors even for quite small values of $\sigma_x/R$ while the
compensation works up to a point where the upper limit of
integration ($2R-\delta$) is felt, and the compensation rapidly
deteriorates as we move beyond this limit. The upper limit on
$\sigma_x(\tau)/R$ defines a critical time lag $\tau_{crit}$ where
measurements for $\tau>\tau_{crit}$ exhibit a serious, systematic
error. For $p=2$ the critical limit is
$\sigma_x(\tau_{crit})/R\sim 0.5$ while it is
$\sigma_x(\tau_{crit})/R\sim 0.4$ for $p=8$. Actually these
estimates are optimistic because $G$ in practice tends to have
fatter tails than a Gaussian. This is shown in
Figure~\ref{fig:Glog}.

In order to get an estimate of $\tau_{crit}$ we set
$\sigma_x(\tau)\sim\sigma_u \tau$ leading to
\begin{equation}
\frac{\tau_{crit,p}}{\tau_\eta}\sim
C_p\,\frac{R}{\sigma_u\tau_\eta}=
15^{-1/2}C_p\frac{R}{L}\,R_\lambda \label{taucritSmallScale}
\end{equation}
with $C_2\sim$0.6, $C_4\sim$0.53, $C_6\sim$0.49 and $C_8\sim$0.45
and the integral scale $L=\sigma^3/\varepsilon$. $L$ was chosen
because it is defined by the geometry of the apparatus used
independent of $R_\lambda$. From (\ref{taucritSmallScale}) we see
that $\tau_{crit}/\tau_\eta$ {\it increases} with $R_\lambda$,
thus making it easier to measure at high Reynolds numbers. In
other respects, such as the demand on frame speed, it of course
gets harder. If we wish to study large time scales, the ratio
$\tau_{crit,p}/T_L$, where $T_L=\sigma^2/\varepsilon$ is the
Lagrangian integral time scale, could be more relevant and we can
note that
\begin{equation}
\frac{\tau_{crit,p}}{T_L}=C_p\frac{R}{L}.
\label{taucritSmallScale2}
\end{equation}
In other words, finite size effects are not affected by the
Reynolds number at large time scales.

Using eqn.~\ref{taucritSmallScale} with the present data we find
$\tau_{crit,2}/\tau_\eta=18$, $\tau_{crit,4}/\tau_\eta=16$,
$\tau_{crit,6}/\tau_\eta=14$ and $\tau_{crit,8}/\tau_\eta=13$.
This is when a 5\% error is accepted. Inspection of the data shows
that at these values the (compensated) integrands in
eqn.~\ref{s2W2} are indeed just covered within $2R-\delta$.
Without compensation we find very small critical limits:
$\tau_{crit,2}/\tau_\eta=3.3$, $\tau_{crit,4}/\tau_\eta=2.0$,
$\tau_{crit,6}/\tau_\eta=1.6$ and $\tau_{crit,8}/\tau_\eta=1.2$.

Other studies have also looked at the bias effect of Lagrangian
statistics~\cite{mordant:2004b,biferale:2008,berg:2005}. The
criterion in eqn.~\ref{taucritSmallScale} and
eqn.~\ref{taucritSmallScale2} are the strictest yet presented in
the literature. The most important lesson is, however, not the
limits suggested by the equations themselves, but the
compensation: without this, even small time lag statistics are
heavily biased as seen in Figure~\ref{ErrorPlots}.

\section{Inertial range scaling} \label{inertialRangeScaling}
\subsection{Presentation of data}
\begin{figure}[h]
\begin{center}
\includegraphics[width=\columnwidth]{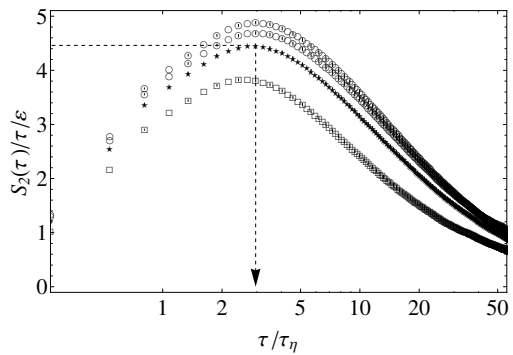}
\caption{{\footnotesize $C_0:$ The circles represent the two
horizontal directions while the squares represents the axial
direction. The stars are the mean (eqn.~\ref{eq:SpDef}). The arrow
indicates the maximum $C_0=4.46$ at $\tau=2.96 \tau_{\eta}$. }}
\label{fig:C0}
\end{center}
\end{figure}

The linear dependence of $Re_{\lambda}$ on $T_L/\tau_{\eta}$
implies that a very high Reynolds number is needed in order to
obtain a clear Lagrangian inertial range. \citet{yeung:2002}
concluded, based on extrapolations from Eulerian fields in DNS,
that at least $Re_{\lambda}\sim 600-700$ is needed. Experimental
flows at $Re_{\lambda}=1000$~\cite{mordant:2004b} and
$Re_{\lambda}= 815$~\cite{ouellette:2006b} do, however, not show a
very pronounced range with a linear regime in the second-order
structure function, $S_2(\tau)=C_0 \varepsilon \tau$, and one
could speculate if such a range exists at all. $C_0$ plays a
crucial role in stochastic models~\citep{sawford:2001} and has
been shown to reflect anisotropy in the large-scale forcing
\citep{ouellette:2006b}. In Figure~\ref{fig:C0} we present results
of $C_0$ for the isotropic ensemble as well as for the three
directions.
According to $S_2(\tau)$, $C_0$ should be determined from a
plateau in the inertial range. The inertial range is almost
vanishing in our experiment. The $\star$ symbols are calculated
from eqn.~\ref{eq:SpDef}. The maximum is $C_0=4.46$ at a time lag
$\tau/\tau_{\eta}=2.96$ and therefore mainly associated with small
scales. The $\circ$ symbols represent the horizontal directions
($C_0=4.88$ and $C_0=4.69$) while the $\Box$ symbols represent the
axial direction ($C_0=3.83$). A rough estimate of the error on
$C_0$ is $0.3$, originating from a $10\%$ error in the
determination of the kinetic energy dissipation $\varepsilon$. The
statistical error is essentially zero.

It is interesting to see that the slight anisotropy in the forcing
is manifested all the way down to $\tau \sim \tau_{\eta}$. The
propellers forcing the flow are attached to four rods placed in
the corners of the tank. The reason for the horizontal components
being different is probably small differences in the vertical
placement of the propellers on the rods. The lack of small-scale
isotropy in the current experiment should not necessarily be taken
as a failure of Kolmogorov's hypothesis of local isotropy. For
such a statement the Reynolds number in our experiment is simply
not high enough. Other experiments at much higher Reynolds number
do, however, all indicate that the large scale inhomogeneities are
also present at smaller scales although with smaller
amplitude~\cite{shen:2000,shen:2002,ouellette:2006b}. These
findings are striking and may suggest that the hypothesis of local
isotropy and the concept of locality should be
reviewed~\cite{tsinober}.

A theory that demands isotropy, as is the case of most K41-like
predictions can literarily not be falsified, since the perfect
experiment with isotropic forcing and hence isotropy on the
smallest scales can not be constructed. This was one of the
motivations  behind the construction of the isotropic ensemble
given in eqn.~\ref{eq:SpDef}.

Alternatively one can calculate $C_0$ from the velocity spectrum.
Arguments put forward by \citet{lien:2002} state the inertial
range scaling is easier to obtain in Fourier space through the
velocity spectrum. However, no difference was observed in the
present data set through such an analysis (not shown).

\begin{figure}[htp]
\begin{center}
\includegraphics[width=\columnwidth]{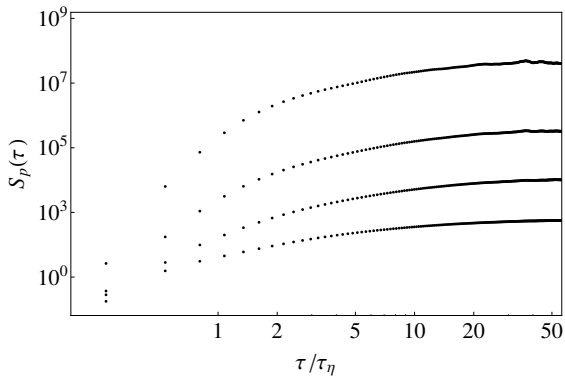}
\caption{{\footnotesize Lagrangian structure functions $S_p(\tau)$
as a function of time lag $\tau$. $p$ is increasing upwards with
$p=2,4,6,8$. The different structure functions have been shifted
vertically for clarity.}} \label{fig:sm}
\end{center}
\end{figure}

We now look at the higher order structure functions. We want to
quantify the degree of intermittency through anomalous scaling
exponents. The structure functions $S_p(\tau)$ for $p=2,4,6,8$ are
displayed in Figure~\ref{fig:sm}. It should be remembered that the
data are heavily influenced by finite volume bias for time lags
$\tau \geq \tau_{crit}$. The most important conclusion to state
from the plot is the evident lack of power law behavior and hence
a $K41$ scaling regime.

Motivated by the lack of a clear inertial range in accessible
turbulent data, \citet{benzi:1993} introduced Extended Self
Similarity (ESS). Even though it was originally applied to
Eulerian data it in can easily be adapted to Lagrangian. Instead
of plotting $S_p(\tau)$ against time lag $\tau$, $S_p(\tau)$ is
plotted against the structure function not affected by
intermittency for the corresponding time lag.

In more general terms we define in the ESS context the anomalous
scaling exponents $\zeta_p(\tau)$ through
\begin{equation}
\zeta_p(\tau) = \frac{d \log{[S_p(\tau)]}}{d \log{[S_2(\tau)]}}.
\end{equation}
It has obvious advantages compared to conventional ad hoc power
law fitting procedures~\cite{arneodo:2008}. It can, however, be
difficult to quantify from experimental data due to the derivative
which is very sensitive to noise.

\begin{figure}[htp]
\begin{center}
\includegraphics[width=\columnwidth]{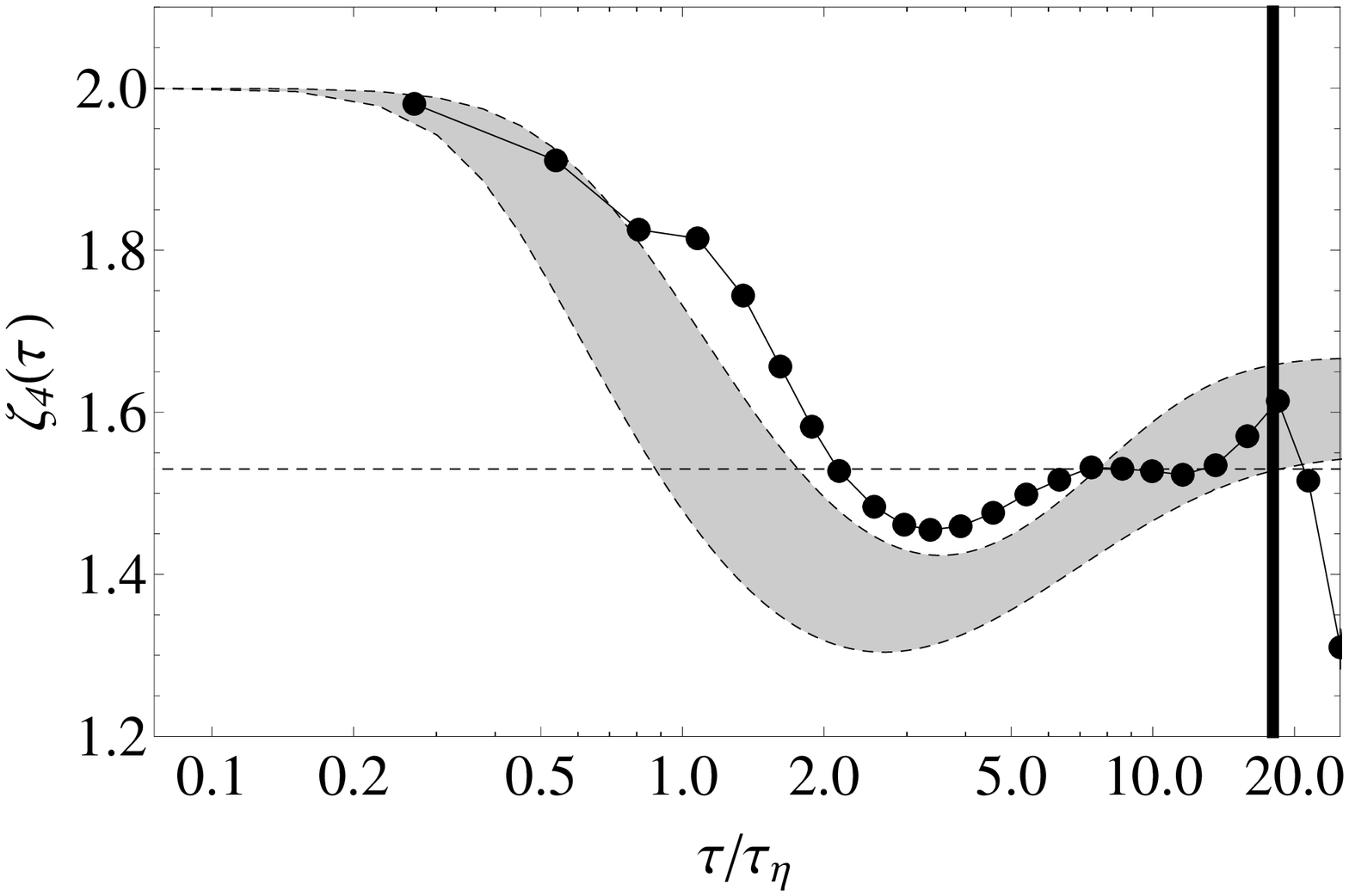}
\includegraphics[width=\columnwidth]{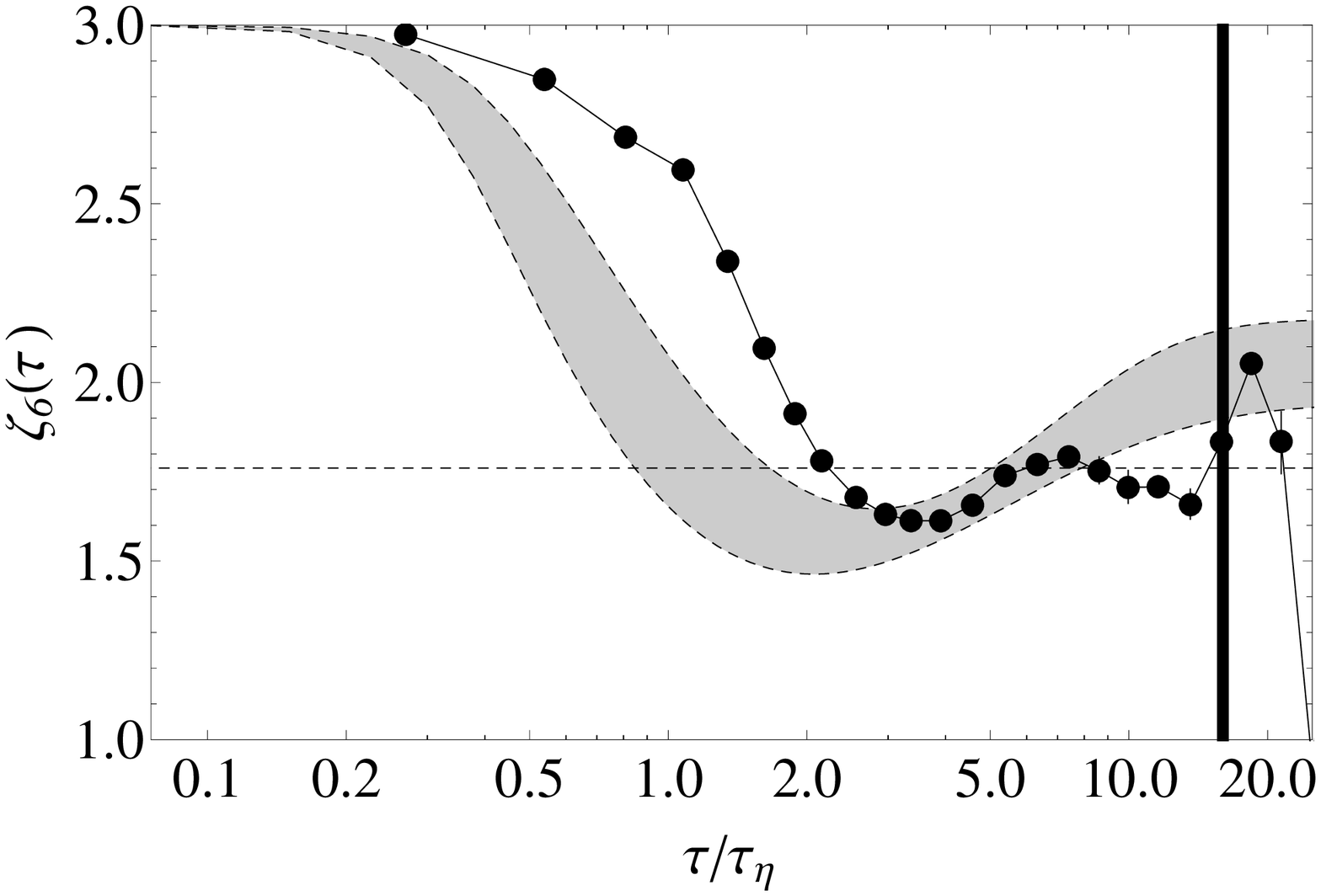}
\includegraphics[width=\columnwidth]{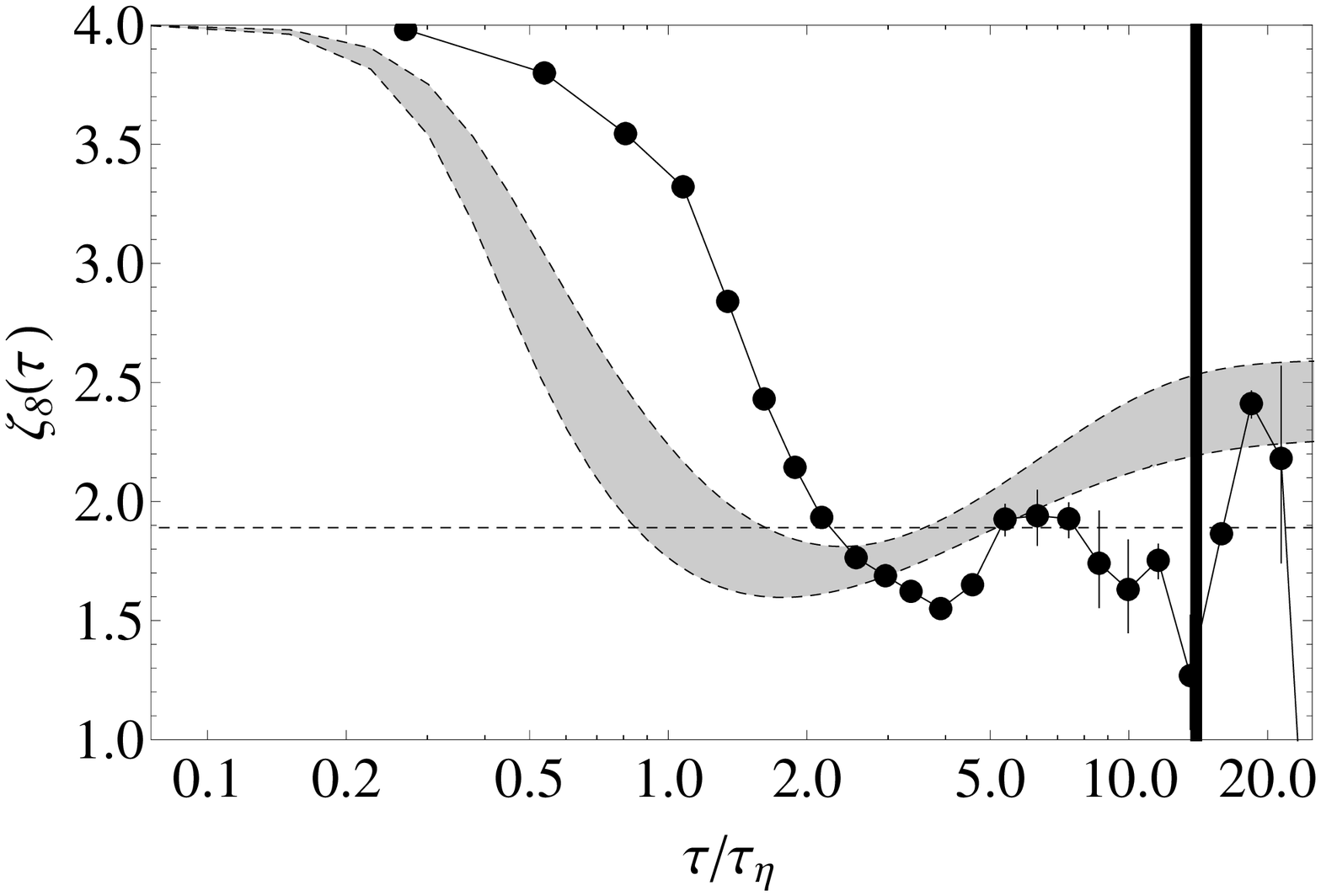}
\caption{{\footnotesize $\zeta_p(\tau)$. {\em Top:} $p=4$, {\em
middle:} $p=6$ and {\em bottom:} $p=8$ (solid curves). The shaded
areas represent the the multifractal predictions: the upper
boundary is based on longitudinal Eulerian structure functions
while the lower boundary is based on transverse Eulerian structure
functions. The thick black vertical lines denote the critical time
lags according to Section~\ref{finiteVolume}. We find the large
time lag saturation levels to be $1.53, 1.76, 1.93$ for $p=4,6,8$.
The error bars refer to statistical errors.}} \label{fig:slopes}
\end{center}
\end{figure}

$\zeta_p(\tau)$ for $p=4,6,8$ are plotted in
Figure~\ref{fig:slopes}. Two general trends are observed in all
three figures.

A dip around $2\tau_{\eta} \leq \tau \leq 5\tau_{\eta}$ is
observed. Hereafter a plateau is reached. For increasing order $p$
the dip becomes larger and $\zeta_p$ saturates at a higher level
in agreement with the findings and speculations by
\citet{arneodo:2008}. The saturation levels are displayed with
horizontal lines at $1.53$, $1.76$ and $1.93$ respectively. At
time lags around $\tau_{crit}$ we see that the error bars grow
significantly at least for $p=8$. With increasing $p$ it even
happens before $\tau_{crit}$ as also suggested in
Section~\ref{finiteVolume} and makes a plateau hard to observe.

\begin{figure}[htp]
\begin{center}
%\hspace*{-2.5cm}
\includegraphics[width=1.0\columnwidth]{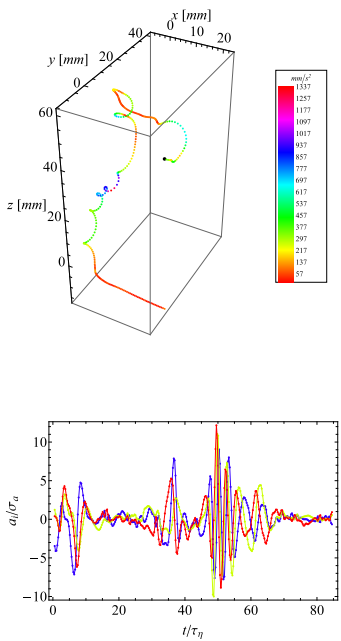}
\caption{{\footnotesize {\em Top} Sample particle. The color
denotes the magnitude of the acceleration. {\em Bottom} Particle
acceleration in units of the standard deviation $\sigma_a$ for all
three components. }} \label{fig:particle}
\end{center}
\end{figure}

The dip has been associated with vortex
trapping~\cite{biferale:2005b} where particles are trapped in
strong vortices with time scale close to $\tau_{\eta}$. An example
of such a particle is displayed in Fig.~\ref{fig:particle}. At
around $t\sim50\tau_{\eta}$ the particle experiences extreme
accelerations (around $10$ times the rms value) and seems to be
caught in a vortex-like structure.

We thus support the findings already reported
in~\citet{biferale:2005b,laporta:2001} in studies of much more
vivid flow. However, we demonstrate that the value of the Reynolds
number is not necessarily crucial in order to observe
characteristic turbulence features in the Lagrangian frame as is
also known from low Reynolds number DNS. Differences in DNS and
physical flows are, however, notable and careful analysis of both
is therefore important in order to obtain a complete picture.

It is also striking that the maximum of $S_2(\tau)/\tau$ is at
$\tau=2.96\tau_{\eta}$ which is very close to the position of the
viscous dip of $\zeta_4(\tau)$. Whether this has any significance
is not at all clear. If so, it is a challenge for stochastic
models, which do not take the viscous dip into account.

\subsection{Multifractal model}
In Fig.~\ref{fig:slopes} we have also plotted the multifractal
prediction. The multifractal model is  developed in the Eulerian
frame by \citet{parisi:1983} to characterize the spatial structure
of dissipation in turbulence it was later adapted in the
Lagrangian frame by \citet{borgas:1993}. It gives a translation
between the two frames and as such works as a bridge between them.
Work presented in
\cite{biferale:2004,biferale:2005b,chevillard:2003,mordant:2002,mordant:2004b,xu:2006b}
has shed light on the issue of multifractals in the Lagrangian
frame through a number of high Reynolds number experiments and DNS
which were well captured by the theory through ESS scaling
relationships for time lags $\tau \geq \tau_{\eta}$. With the
extension by \citet{chevillard:2003} the multifractal framework is
now capable of taking into account also the vortex trapping
behavior taking place at time lags $\tau \sim \tau_{\eta}$.

In \cite{arneodo:2008} it was shown for $\zeta_4(\tau)$ how the
multifractal model matches results from experimental and numerical
data on all time scales independently of Reynolds number. The data
set used in this paper is the one denoted EXP 1 in that paper.
Here we show that for $p=6$ and $p=8$ the multifractal model seems
somewhat less perfect.

In the multifractal model the flow is assumed to possess a range
of scaling exponents $h_{min},...,h_{max}$ with a certain
probability so that the velocity difference by separation $r$ is
$\delta_r u \sim r^h$. For each scaling exponent $h$ there is a
fractal set with a $h$-dependent dimension $D(h)$. The embedding
dimension is three $(r\in\mathcal{R}^3)$ and hence $D(h)\leq3$ for
all $h$. The probability $P_h(r)$ of having an exponent $h$ at
separation $r$ is therefore proportional to $3-D(h)$. From
dimensional arguments the Eulerian velocity fluctuation $\delta_r
u$ is related to a Lagrangian time lag $\tau$. That is $\tau\sim
r/\delta_r u$. Recently focus on this relation has cast serious
doubt on its usage~\cite{homann:2007,yakhot:2008,kamps:2008}. In
\cite{kamps:2008} it is shown how the relation $\tau\sim
r/\delta_r u$ is the limiting case of something more general. In
three dimensional turbulence they show that Lagrangian statistics
is much more influenced by Eulerian integral and dissipation
scales than the simple picture suggest, where the time lag $\tau$
is only associated with eddies of size $r$. This is also what we
already know from Figure~\ref{fig:G} and from eqn.~\ref{eq:greens}
where contributions from all scales is included in the integral.
The conclusion must be that good statistics of all Eulerian scales
are necessary in order to calculate Lagrangian structure function
- even at small time lags. At present DNS does not resolve the
statistics of the largest scales sufficiently. PTV experiments do
not have this problem.

Following \cite{arneodo:2008} closely we can calculate the
multifractal prediction for $p=4,6,8$. We choose the same model
constants and functional form of $D(h)$ as in \cite{arneodo:2008}
since these were shown to fit a large number of experiments and
DNS simulations for $p=4$.

We again take a look at Figure~\ref{fig:slopes}. The result in
\cite{arneodo:2008} ($p=4$) is reproduced: Following the
multifractal longitudinal curve (upper) in the dip quite close,
$\zeta_4(\tau)$ saturates at a value close to the multifractal
transverse curve (lower). For $p=6$ and $p=8$ the multifractal
predicted curves does not fit the data. First, the plateau in the
data does not reach the level of the multifractal prediction
before finite volume bias effects become important. This bias has
the effect of lowering the value of $\zeta_p(\tau)$. Second, for
both moments the dip is shifted towards the left. The dip minimum
does, however, seem to match the data. In the multifractal model
there is a free parameter included in the model definition of the
dissipative time scale. We have set this constant, $t_{scale}$
equal to $7$. Its only job is to scale the $\tau/\tau_{\eta}$
axis. The same constant value was used in \cite{arneodo:2008} in
order to fit the model prediction to data for $p=4$. Since
$t_{scale}$ must be independent of $p$, we could equally well have
fitted $t_{scale}$ to the data for $p=6$ or $p=8$. If we had done
so we would encounter a bad match for $p=4$ and $p=8$ or $p=4$ and
$p=6$ respectively.

\begin{figure}[htp]
\begin{center}
\includegraphics[width=1\columnwidth]{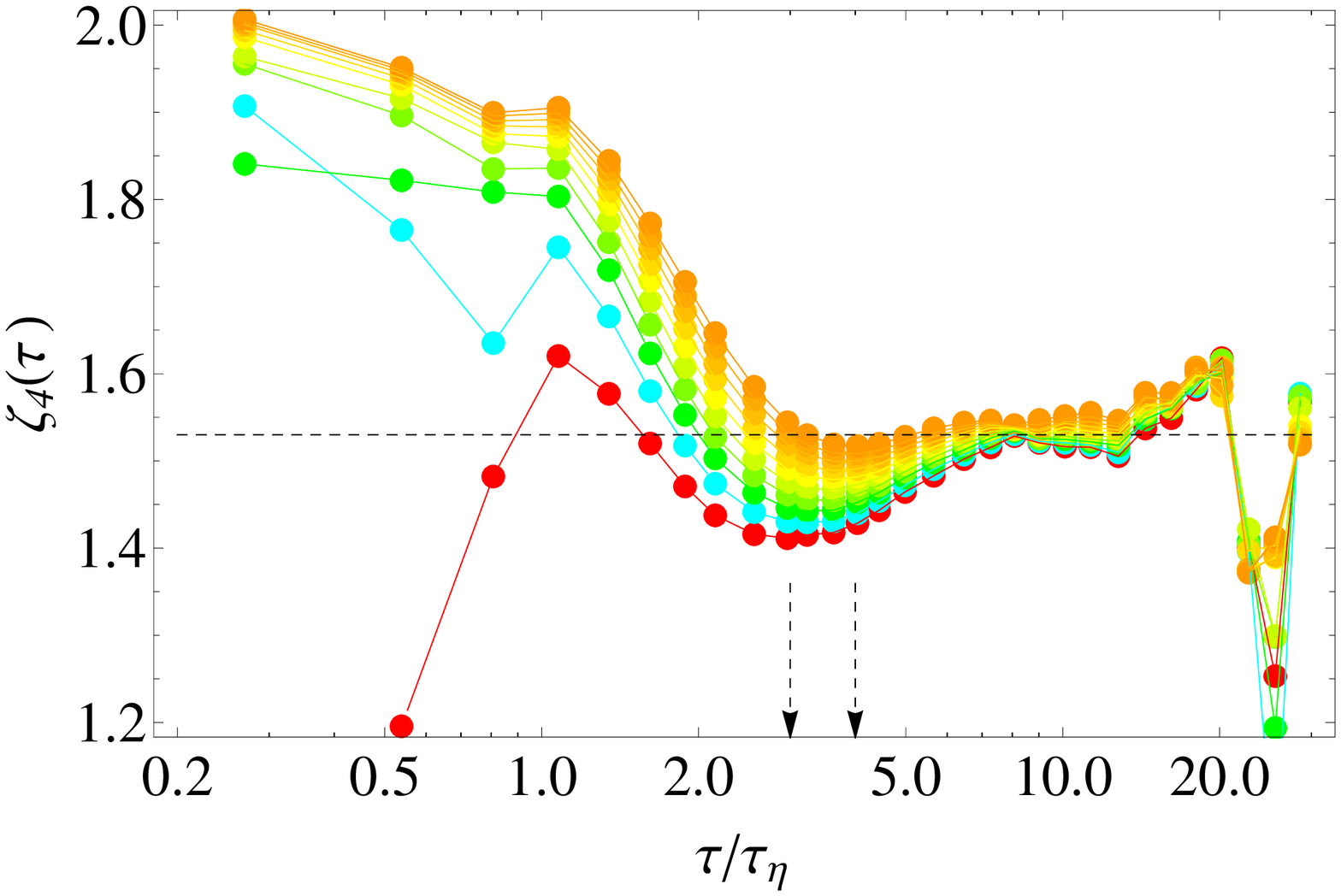}
\includegraphics[width=1\columnwidth]{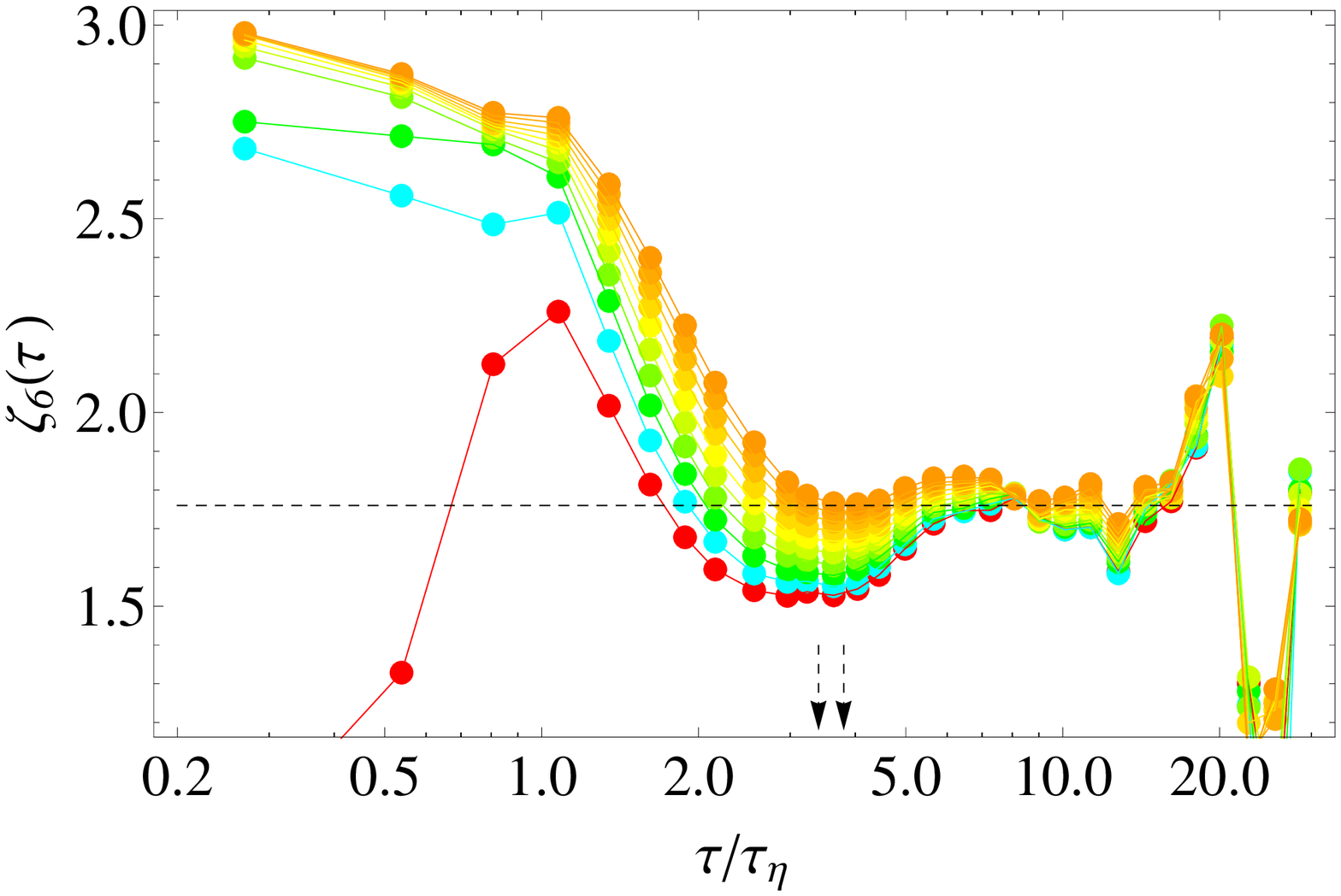}
\includegraphics[width=1\columnwidth]{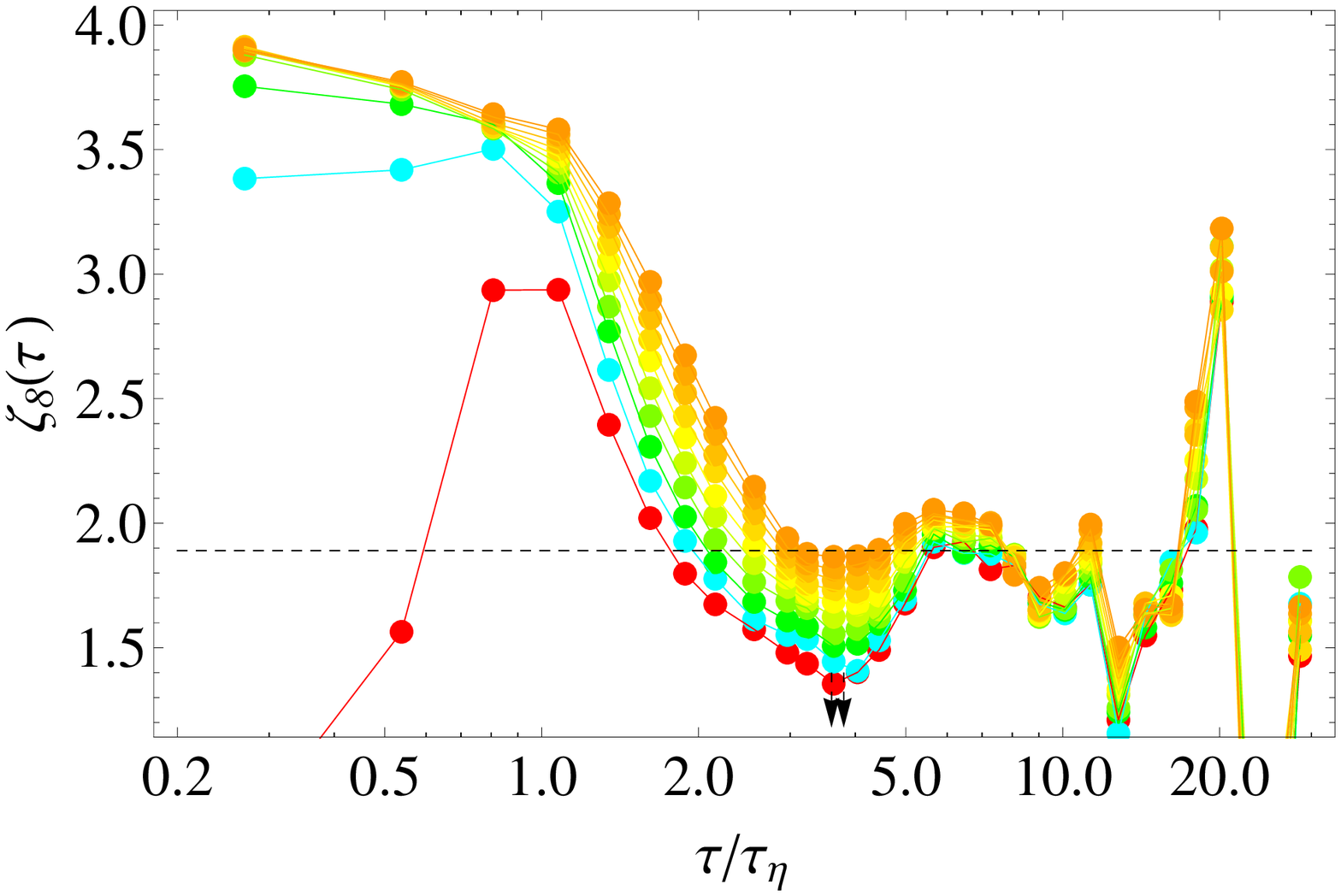}
\caption{{\footnotesize $\zeta_p(\tau)$. {\em Top:} $p=4$, {\em
middle:} $p=6$ and {\em bottom:} $p=8$ (solid curves). The colors
denote different filter $N$ (increasing upwards from $N=1$ (red)).
Arrows indicate minimum position for $N=1$ (left arrow) and $N=10$
(right arrow).}} \label{fig:filterSlopes}
\end{center}
\end{figure}

Why the multifractal model fail to predict the different moments
of $\zeta_p(\tau)$ is a relevant question. We could speculate that
the temporal resolution in our experiment is not high enough to
resolve the smallest scale. In order to investigate this we look
at the importance of filtering. We picture the increase of filter
length as a way of decreasing the temporal resolution and
calculate $\zeta_p(\tau))$ as a function of filter length, $N$.
For $N\leq10$ the resuls are presented in
Figure~\ref{fig:filterSlopes}.
With increasing filter length we see that the dip region is
depleted and the position of the minimum on the $\tau$ axis is
shifted to the right (the arrows indicate the position of the
local minimum for $N=1$ and $N=10$). The shift in position is
decreasing with increasing $p$ and hardly observable for $p=8$.
The saturation level is more or less constant with $N$. Both the
quantitative and qualitative shape of the dip region of
$\zeta_p(\tau)$ are therefore quite sensitive to the filter length
$N$. How does this relate to the multifractal model? In
Figure~\ref{fig:slopes} we see that the dip position of the
multifractal model is almost constant with $p$, while the data
show a shift towards smaller time lags when $p$ is increased. On
the other hand, we observe in Figure~\ref{fig:filterSlopes} that
for $p=4$ the shift is towards larger time lags whereas it is
almost non-existing for $p=8$. That is, for a higher resolution,
here pictured as a lower filter, the multifractal model would work
even worse.

\section{Discussion and conclusions}
Analyzing Lagrangian data uncritically can lead to substantial
biases. Building on the ideas first presented by~\citet{ott:2005}
we have found a robust way to compensate for the finite volume
bias. We show that the time lags for which finite volume bias can
be neglected are limited. We also saw that the finite volume
effects increase for increasing order $p$ of the structure
functions. Observing extreme statistics might therefore be very
difficult. We are convinced that the present study and its
consequences should be kept in mind when designing future
experiments for measuring Lagrangian statistics: in a high
Reynolds number flow the separation between the integral scale and
the Kolmogorov length is very large. In order to follow particles
and measure the Lagrangian structure functions for large time lags
without finite volume bias the camera chip needs to be extremely
big, which is currently a technical obstacle. Alternative setups
where different camera systems focus on the small and the large
scales simultaneous could be the solution to this problem. DNS
does not suffer from finite volume bias. On the other hand DNS may
still have problems with small time lags due to interpolation from
the Eulerian flow field~\cite{biferale:2008}.

In particular we have studied one-particle statistics in terms of
Lagrangian structure functions. We have looked at the small time
scale behavior which seems to be affected by the large-scale
inhomogeneities present in our flow. This led us to work with
isotropic ensembles. In this way we can test our data against
theories developed in an isotropic frame such as the multifractal
model.

We do not observe any signs of an inertial range and K41 scaling
but by Extended Self-Similarity we are able to extract a
quantitative measure of the structure functions of high order.
From the local slopes of these we calculate the Lagrangian
anomalous scaling exponents and find excellent agreement with
already published results.

Measured local slopes of the Lagrangian structure functions are
quite similar to results obtained with the multifractal model for
$p=4$. With the assumptions and physical reasoning leading to the
development of the multifractal model in mind, this is actually a
bit surprising. Many of the crucial assumptions behind the
multifractal model in the Lagrangian frame are not fulfilled.
First of all the multifractal model is motivated by the invariance
of the Navier-Stokes equation to an infinite number of scaling
groups in the limit of infinite Reynolds number far from present
in our data. Second, an exact result such as the 4/5 law does not
exist in the Lagrangian frame. The phenomenological picture of the
multifractal model with the flow region consisting of {\em active}
and {\em inactive} regions in direct contrast to the Richardson
picture where eddies are {\em space filling} does, however, fit
observed flow features such as a dip region in $\zeta_p(\tau)$ and
anomalous scaling.

For $\zeta_6(\tau)$ and $\zeta_8(\tau)$ the situation is
different. The multifractal model fails to describe the data:
although the qualitative behavior is similar to $\zeta_4(\tau)$
the shift in dip position as a function of $p$ is much more
pronounced in our data than in the multifractal model. We are
curious to see results from other experiments and DNS simulations
for $p>4$. The simple bridging relation, where a time lag is
solely associated to the time scale, $\tau$ for the local eddy of
size $r$, assumed in the multifractal model might be too crude.
Also the filtering was shown to have an effect on the dip region:
the strength for all orders of $p$ while the position was only
dependent on filtering up to $p=6$. Since almost all data sets,
both from physical experiments and from DNS, are noisy and hence
must be filtered, conclusions on $\zeta_p(\tau)$ should therefore
be made with great care.

\acknowledgments J. Berg is grateful to L. Biferale and two
anonymous reviewers for valuable comments and suggestions.

%@Preamble{{\input{apalike.sty}}}
\end{document}